\documentstyle[11pt]{article}
\begin{document}

\begin{centering}
{\bf Entropic Forces in Binary Hard Sphere Mixtures: Theory and Simulation} \\
\vspace{1em}

Ronald Dickman$^{\dagger,a}$, Phil Attard$^{\ddagger,b}$, 
and Veronika Simonian$^{\dagger,c}$ \\
\vspace{1em}

$^{\dagger}$Department of Physics and Astronomy, Lehman College, CUNY,
Bronx, NY 10468-1589, USA \\and \\
$^{\ddagger}$School of Chemistry, F11, University of Sydney, 
NSW, 2006 Australia       \\             
\vspace{1em}

\begin{abstract}
We perform extensive Monte Carlo simulations of binary hard-sphere mixtures
(with diameter ratios of 5 and 10),
to determine the entropic force between (1) a macrosphere and a hard wall, and (2) a
pair of macrospheres.  The microsphere background fluid (at volume fractions ranging from
0.1 to 0.34) induces an entropic force on the macrosphere(s); the latter component
is at infinite dilution.  We find good overall agreement, in both cases, with the predictions 
of an HNC-based theory for the entropic force.  Our results also argue for the validity of the Derjaguin approximation relating the force between convex bodies to that between
planar surfaces.  The earlier Asakura-Oosawa theory, based
on a simple geometric argument, is only accurate in the low-density limit.
\vspace {0.3truecm}

\end{abstract}
\end{centering}
\vspace{1.0truecm}

\noindent $^a${\small e-mail address: dickman@lcvax.lehman.cuny.edu  }\\
$^b${\small e-mail address: attard@chem.usyd.edu.au}\\
$^c${\small e-mail address: vxs@lcvax.lehman.cuny.edu}
 
\newpage

\noindent {\bf I Introduction}
\vspace{1.0truecm}

Entropic forces in colloidal suspensions and in polymer-colloid systems are of longstanding
and continuing interest \cite{asak54,RUSSEL,jldg79,vrij76,ghr86,yhd92,dy94}.  Recent
experiments have probed the phase diagram of binary colloidal suspensions \cite{KAPLAN94}, 
and have determined the entropic potential between a colloidal particle and a wall,
induced by a smaller colloid component \cite{KAPLAN94a,DINSMORE}.   
While the first steps in  the theory of entropic interactions were
 taken some decades ago, in the geometrical arguments of Asakura
and Oosawa \cite{asak54}, and the extension of Percus-Yevick 
theory to hard-sphere mixtures
\cite{LEBOWITZ}, substantial refinements in integral-equation based 
approaches to the problem
have been proposed only recently \cite{Attard90}.  At present the most 
reliable predictions are
those derived using the hypernetted chain (HNC) equation, corrected 
by including  bridge
diagrams up to third order in density, yielding the so-called HNCP theory.
Recent experiments
and renewed theoretical activity motivate our study of entropic forces in simulations of
the simplest pertinent model --- a binary hard-sphere fluid --- in hopes of providing a
critical test of theory, and of deciding whether the hard-sphere model, and current theoretical
approaches for the latter, are adequate for a detailed understanding of the experimental 
results.

In {\em athermal} systems (in which all allowed configurations 
have the same energy), entropic interactions alone determine 
any structure at interparticle separations beyond the range 
of the (hard-core) potential.  In the 
hard-sphere fluid (the prime example of an athermal model), 
each molecule is surrounded by a sphere of radius equal to the 
molecular diameter, from which the centers of other molecules 
are excluded.  Since the overlap of exclusion spheres associated with any two
molecules increases the available space for the remaining molecules, maximization 
of entropy favors small separations between nearby particles, that is, a peak
in the radial distribution function, $g(r)$, at contact.  This line of 
reasoning forms the basis for the Asakura-Oosawa (AO) theory.  
While the latter assumes ideality of the microsphere component and yields a 
purely attractive entropic force, excluded volume considerations suggest
a repulsive force for macrosphere separations on the order of the microsphere 
diameter. To go beyond simple geometric arguments requires a detailed
theory for the structure of a binary fluid, since the entropic 
interaction (or potential of mean force) is obtained from the 
interspecies two-point distribution function.  
To test the AO and HNCP predictions, we perform extensive Monte Carlo 
simulations of a system consisting of one or two hard `macrospheres' in a fluid of hard `microspheres.'  (The diameters of the two species have a ratio of 5 or 10.)

In this paper the HNCP approximation is tested for the first time against
Monte Carlo simulations for interacting macrospheres in a hard-sphere solvent.
Previously this approximation has been tested for a pure hard-sphere solvent 
\cite{Attard90}, and also for interacting planar walls
in a hard-sphere fluid \cite{Attard90,Attard91}. 
The latter test relied upon the validity of the Derjaguin approximation
\cite{Derjaguin34,White83,Attard92}, 
which relates the force between convex bodies to a geometrical
factor times the interaction free energy of planar walls.

This paper also tests the Asakura-Oosawa depletion-attraction theory,
which predicts an adhesion between macrospheres due to exclusion of 
microspheres from the region between them. 
The solvent-mediated force between two hard solutes 
can be expressed formally in terms of the contact density \cite{Attard89},
and the AO expression approximates the latter as the
bulk density away from the contact region.
It can therefore be expected to be valid for low solvent densities,
but again its precise regime of validity and its dependence upon
solute diameter remains to be tested.
There is some evidence that the Asakura-Oosawa expression
is accurate for the contact adhesion between
hard-sphere solutes in hard-sphere solvents,
but these tests were carried out with 
the inhomogeneous Percus-Yevick results in superposition approximation 
\cite{Attard89}.

Our simulations show that the entropic force can be repulsive as well as
attractive, as noted recently for the force between planes immersed
in a fluid \cite{MAO}.  Our force profiles, which are attractive near
contact, exhibit a repulsive peak at a separation of about one microsphere
diameter, and show strongly damped oscillations at larger separations.
The details of this oscillatory structure are reproduced quite
faithfully by the HNC theory, but are entirely absent from the AO theory, which
predicts a purely attractive potential with a range of one microsphere diameter.

The balance of this paper is organized as follows.  In Sec. II we review 
AO theory and present an intuitive argument for the repulsive barrier, 
and then outline the HNCP approach.  The simulation method is
described in Sec. III, with results and analysis following in Sec. IV.  We 
summarize our conclusions in Sec. V.
\vspace{1em}

\noindent {\bf II Theory}
\vspace{1em}

{\bf IIa Elementary Geometrical Argument}

We begin by reviewing the geometrical argument of Asakura and Oosawa \cite{asak54} for
the entropic force between a pair hard spheres, of radius $R$, their 
centers separated a distance $2R + D$, and immersed in a fluid
of particles with hard-sphere radius $r$.
If we treat the fluid as an ideal gas of $N$ particles, then its Helmholtz free energy, to within terms independent of $R$, $r$, and $D$, is
\begin{equation}
{\cal F} = -N k_B T \ln V' \;,
\label{hfe}
\end{equation}
where $k_B$ is Boltzmann's constant, $T$ is temperature, and $V'$ is the volume
available to the fluid particles.  Since the particles are prohibited from the
{\em exclusion spheres} of radius
$R+r$ about the large spheres, the available volume is
\begin{equation}
V' = V - \frac{8 \pi}{3} (R+r)^3 + v_{ov} \;,
\label{vavail}
\end{equation}
where $V$ is the system volume and $v_{ov}$ is the {\em overlap volume} of the two
exclusion spheres.  The entropic force between the two spheres is therefore
\begin{equation}
F = -\frac{\partial {\cal F}}{\partial D} = \frac{Nk_B T}{V'}  \frac{\partial v_{ov}}{\partial D} \;.
\label{force1}
\end{equation}
Since $\partial v_{ov}/\partial D$ is just the projected area of intersection of the two
exclusion spheres, simple geometry yields
\begin{equation}
F_{ss,AO} = -\rho k_B T \pi (r - \frac{D}{2})(2R + r + \frac{D}{2}) \;,
\label{forceao}
\end{equation}
for $D \leq 2r$, and zero for larger separations ($\rho = N/V$ is the fluid density).

While this argument invokes an ideal gas assumption 
that is unjustified at significant densities, and applies it 
inconsistently (since the macrospheres exclude particles
from a region of radius $R+r$ not $R$), it does provide a useful 
estimate of the force.  The associated sphere-sphere entropic potential is
\begin{equation}
V_{ss,AO} = - \pi \rho k_B T  (2r -D) \left[r \left (R+\frac{2r}{3} \right) - 
\frac{D}{2} \left( R+\frac{r}{3}  \right) -\frac{D^2}{12} \right] \;,
\label{potao}
\end{equation}
for $D \leq 2r$, and zero for larger separations.  For a macrosphere of radius $R$ 
centered at a distance $R+D$ from a hard wall, a similar argument leads to
\begin{equation}
F_{ws,AO} = -\rho k_B T \pi (2r - D)(2R + D) \;,
\label{foraow}
\end{equation}
and
\begin{equation}
V_{ws,AO} = - \pi \rho k_B T  (2r -D)\left[2r \left( R+\frac{r}{3} \right) - 
D \left( R-\frac{r}{3} \right) -\frac{D^2}{3} \right] \;,
\label{potaow}
\end{equation}

\noindent for $D \leq 2r$, and again zero for larger separations.
Eq. (\ref{foraow}) yields a force at contact of 
$F_{ws,AO}(D\!=\!0) = -4 \pi r \rho k_B T R$, and since we use $r=1/2$ in the
simulations, this motivates our defining a scaled force 
\begin{equation}
f^*_{ws} (D) \equiv \frac {F(D)}{2\pi R \rho k_B T} \; ,
\label{fscws}
\end{equation}
for the wall-sphere case.
Similarly, in the two-sphere case, reference to Eq.(\ref{forceao}) suggests defining  
\begin{equation}
f^*_{ss} (D) \equiv \frac {F(D)}{\pi R \rho k_B T} \; .
\label{fscss}
\end{equation}
Normalizing $F(0)$ to $R$ renders it independent of $R$ in AO theory,
and in the Derjaguin approximation as well (see below).  (At contact,
AO theory obeys the
same scaling as the Derjaguin approximation; the scaling
is only approximate for larger separations.)
If Asakura-Oosawa theory were exact, we would have $f^*_{ws} (0) =  -1$,  
and $f^*_{ss} (0) = -(1+\frac{1}{4R})$.
Despite the approximations involved in AO theory, these
expressions provide a useful basis for comparing results for spheres of different sizes.
It is also useful to note that the expressions for forces and potentials may be cast in dimensionless form if we use the volume fraction $\eta \equiv 4 \pi r^3 \rho /3 $ and the diameter ratio $\xi \equiv R/r$.  The wall-sphere potential, Eq. (\ref{potaow}), 
at contact is then given by
\begin{equation}
\frac {V_{ws,AO}(0)}{k_B T}  = - 3 \eta \xi \left( 1 + \frac{1}{6\xi} \right) \;.
\label{pndaow}
\end{equation}

To go beyond simple geometric arguments, one must determine the
background-fluid density at contact with the particle.  It is
straightforward to show that the force on a sphere immersed 
in a fluid with local density $\rho ({\bf r})$ is \cite{Attard89}
\begin{equation}
{\bf F} = - k_B T  \int _S \rho \; \hat{{\bf n}} dA     \;,
\label{forint}
\end{equation}
where the integral runs over the surface of a sphere of radius $R+r$,
centered on the spherical particle, and $\hat{{\bf n}}$ is the outward normal unit
vector.  For the symmetrical arrangements considered here only $F_x$
is nonzero, given by
\begin{equation}
F_x = - 2 \pi k_B T  (R+r)^2 \int  \rho(R+r,\theta)  \cos \theta \sin \theta d\theta     \;.
\label{forintx}
\end{equation}
This expression forms the basis of our force calculation in simulations.

If the microsphere fluid were an ideal gas, the entropic interaction
would be the purely attractive one predicted by AO theory.  In fact, the contact
density is considerably in excess of $\rho _{bulk}$.  (At a planar wall the
contact density is simply $p/k_B T$, where $p$ is the pressure, and the contact
density at an isolated macrosphere will approach this value for large $\xi$.) 
The elevated density at a hard surface may again be seen as ensuing from
overlap of two exclusion regions, one associated with a microsphere, the other
with the obstacle, be it a hard wall or a macrosphere.
Thus one might expect the entropic force to grow in proportion to the bulk
{\em pressure} rather than the bulk density.  When a pair of macrospheres is at or near 
contact, however, an additional
compensating factor arises: the contact density is further enhanced in the
vicinity of the corner or channel between the macrospheres.  Here the exclusion
region of a microsphere overlaps the exclusion zones of {\em both} macrospheres.

Consider a pair of macrospheres at contact, with centers along the
$x$-axis at $\pm R$.  If we measure $\theta$ from the positive $x$-axis,
then the contact density vanishes for $\theta > \theta_{max} $, where
$\cos \theta_{max} = - R/(R+r)$.  Indeed, setting the contact density $\rho (R+r,\theta)$
equal to the bulk density $\rho$ for $\theta \leq \theta_{max}$, and to zero for larger
$\theta$, Eq.(\ref{forintx}) yields the AO contact value  $F_x = -\rho k_{B} T \pi r(2R +r)$.
AO theory underestimates the contact density, since
for $R>>r$, $\rho (R+r,\theta \!=\! 0) \simeq p/k_{B}T $.  We have, moreover, 
just argued that the contact density 
{\em increases} as $\theta $ approaches $\theta_{max}$,
yielding a repulsive contribution to $F_x$.  This situation persists
as the macrospheres are separated.  As $D$ approaches the microsphere diameter
$2r$, $\theta_{max} \! \rightarrow \! \pi$, and if the contact density were uniform on 
$[0,\theta_{max}]$, $F_x$ would vanish at $D=2r$.  But since $\rho (R+r,\theta)$
actually increases as $\theta \rightarrow \theta_{max}$,
we expect $F_x$  to vanish at some separation 
$D_0 < 2r$, and to be {\em repulsive} for 
$D \stackrel {>}{\sim} D_0$.  Similar arguments apply in the wall-sphere case.
To summarize, a qualitative consideration of microsphere excluded-volume effects
suggests that (1) the entropic force grows faster than the bulk fluid density, though
perhaps not as rapidly as the bulk pressure, and (2) the entropic force should be
repulsive for $D \approx 2r$.
\vspace{1em}

{\bf IIb Hypernetted Chain Theory} 
\vspace{1em}
  
Hypernetted chain calculations were performed for 
a hard-sphere solvent that included hard-macrospheres
at infinite dilution (singlet method).
Bridge functions were included via a Pad\'{e} approximant constructed 
from the two bridge functions of second and third order in density,
(i.e., it includes all $f$-bond bridge diagrams with two and three density 
field points), as described by Attard and Patey,
and termed by them the HNCP approximation \cite{Attard90}.
This was done for solvent-solvent, solute-solvent, and solute-solute
bridge functions. Thus the computed potentials of mean force
are exact through third order in density,
in contrast to the bare hypernetted chain approximation,
which is exact only through first order.
The reason for using this many bridge diagrams 
is that the accuracy of a given singlet closure decreases by one power
of density for each solute \cite{Attard90,Attard89},
and so for reliable results one needs a sophisticated closure
such as that used here.
The hypernetted chain calculations were performed for two
interacting macrospheres of radius $R=$ 5, 10, and 20 times
that of the solvent hard-spheres.

An alternative closure that could have been used
is the Percus-Yevick approximation.
However for large solutes the singlet method 
(i.e., the Percus-Yevick approximation applied to an asymmetric mixture),
gives a solute-solute radial distribution function that is negative in
places, and the contact values, though analytic, become markedly less 
accurate with increasing diameter ratio \cite{Attard89}.
The spherically inhomogeneous Percus-Yevick approach \cite{Attard89},
which solves the Ornstein-Zernike equation in the presence of a fixed
macrosphere, gives extremely good results for a single solute, 
but it can only be applied to the problem of two interacting solutes 
by invoking a superposition approximation \cite{Attard89}.
While partial tests of the latter suggest it is reasonable in the case
of hard-macrospheres in a hard-sphere solvent at low to moderate
densities \cite{Attard89}, 
the present singlet hypernetted chain approach with bridge
functions is more convenient and has been shown to remain
accurate over the whole fluid regime  \cite{Attard90}. 

Our calculation of the wall-sphere interaction employs the Derjaguin 
approximation \cite{Derjaguin34,White83}, 
which relates the force between convex bodies to a geometrical
factor times the interaction free energy of planar walls.
Specifically,
the interaction free energy per unit area between planar walls
equals the net force between two macrospheres divided by $\pi R$,
which equals the net force between a macrosphere and a planar wall
divided by $2\pi R$.  In other words, $f^*_{ws} = f^*_{ss}$, in our notation.) 
It has been shown that the Derjaguin approximation 
is the exact limiting form for the force
in the asymptotic limit of vanishing curvature
\cite{Attard91,Attard92}, but whether it can be applied to finite-sized
solutes, and at what separations, is not clear.
Previous tests of the Derjaguin approximation 
for hard macrospheres in hard-sphere fluids have been 
in the context of the HNCP approximation \cite{Attard90}
and the inhomogeneous Percus-Yevick approximation \cite{Attard89}.
In this paper the Derjaguin approximation will be tested
directly against simulations of interacting macrospheres
and of a macrosphere interacting with a planar wall.
\vspace{1em}

{\bf III Simulation Method}
\vspace{1em}

We consider a simple model of the colloid mixtures studied in 
recent experiments: a binary hard-sphere fluid with the macrosphere 
component effectively at infinite dilution. In studies of wall-sphere 
interactions, the system is a fluid of unit-diameter hard spheres, 
in a cell with hard walls at $x\!=\!0$
and $x\!=\!H$ (the centers of the spheres are restricted to $0 \leq x \leq H$), and periodic boundaries, with repeat distance $L$, in the $y$ and $z$ directions.  There is a single macrosphere of radius $R$ with its center a distance $R\!+\!D$ from the wall at $x\!=\!0$.  For
cells large enough to render finite-size effects inconsequential, the force on the sphere is a function of the diameter ratio $\xi \!=\! 2R$, the
separation $D$, and the small-sphere volume fraction $\eta$ {\em in bulk}.  (In simulations the latter is not known {\em a priori} but must be determined from the density profile $\rho(x)$.)

In the two-sphere studies, the fluid is placed in a cell periodic in all three
directions.  The two macrospheres, again of radius $R$, 
have their centers a distance $2R \!+\! D$ apart.  The cell dimensions --- $H$ 
along the $x$ direction, $L$ in the perpendicular directions --- are 
large enough that the density profile has a bulk-like plateau in the 
region away from the spheres.

The primary goal of the simulations is to evaluate the force on a
macrosphere using Eq. (\ref{forintx}).  To this end we sample configurations 
of the microsphere
fluid (in the canonical ensemble), with the macrosphere(s) 
{\em fixed}, at separation $D$.
Thus each step in the simulation involves a trial displacement of a
randomly selected microsphere; the new position is accepted as long as
it does not result in an overlap with another microsphere, the macrosphere(s),
or a wall.  Cell-occupancy lists are maintained to streamline testing for
overlap with the other solvent particles. Three copies the system,   
maintained at bulk volume fractions $\eta $ = 0.1, 0.2 and 0.3 are simulated in a 
series of runs at a given $D$.  To reduce the cpu time we use the same sequence 
of random numbers for each copy.  By adding
or removing particles ({\em before} taking any data), we maintain the volume
fraction to within about 0.5\%.
(In the $R\!=\!5$ studies we used approximately 920, 1870, and 2850
microspheres for $\eta = 0.1$, 0.2, and 0.3, respectively.)
In the two-sphere studies, somewhat larger volume fractions
--- $\eta$ = 0.116, 0.229, and 0.341 --- were employed.
For the $R\!=\!5$ wall-sphere studies the cell dimensions were $H=22$, $L=16$;
for $R=2.5$ the corresponding dimensions were 20 and 14.  (All lengths are
in units of the microsphere diameter.)  The two-sphere studies employed cell
dimensions $H=22$ and $L=16$ ($R=5$), and $H=18$, $L=12$ ($R=2.5$)
Fig. 1 shows the density profile $\rho (x)$ of the microspheres 
along the direction perpendicular to
the walls, and illustrates 
the familiar oscillations near the walls, a region of reduced density 
in the vicinity of the macrosphere, and a broad region of constant 
density, representing bulk fluid.  The bulk density $\rho _b$ is figured from 
the profile in this oscillation-free region.

Let $s_{\rm i}$ denote the center-to-center distance 
of microsphere i from the macrosphere.
According to Eq. (\ref{forintx}), calculating the force requires that we know
the the contact density (i.e., the density of 
microspheres with $s=R+1/2$), as a function of the 
polar angle $\theta$.  We follow the usual practice 
of obtaining contact densities by extrapoltaing data 
{\em near} contact.  We avoid a massive data 
storage and extrapolation task by applying this procedure 
not to $\rho (s,\theta)$, but rather to the integral
\begin{equation}
I(s) =  \int  \rho(s,\theta,\phi)  \cos \theta d\Omega     \;.
\label{iofs}
\end{equation}
In practice we divide the space around the macrosphere into shells of thickness
0.02, and determine $I_{\rm i} \equiv \langle \sum \cos \theta_{\rm i} \rangle$, 
where the sum is over all particles in shell i and the brackets 
denote a thermal average.  Shell 1 is then centered a distance of 0.01
from contact, permitting the contact value of $I(s)$ to be determined by
fitting the data near contact with a quartic or lower degree polynomial, 
as illustrated in Fig. 2.  

As one varies the position of the
macrosphere, there are small variations in the bulk density, even after
attempting to compensate for this by adding or removing particles.  In order
to have a set of data representing the force profile at a given
volume fraction, we use quadratic interpolation to obtain 
the force at the fractions stated above.  The resulting correction to the raw data is
generally less than 1\%, smaller in many cases than the 
statistical uncertainty in the measured force.  We estimate 
the latter from the standard deviation over
3 -5 successive runs (each involving $2 \times 10^9$ trial displacements).  
At contact, the relative uncertainty is small --- about 0.5-2 \%.
The absolute uncertainty is roughly independent of separation, and at separations 
of 2-2.5 represents a substantial fraction of the (now quite weak) force.  
This is not a major shortcoming, since at the densities studied, the force is 
nearly zero for $D > 2$.  For comparison with theory and experiment, it is useful to 
compute the entropic {\em potential},
\begin{equation}
V(D) =  \int _D ^{D_{max}} F(\lambda) d\lambda     \;,
\label{entpot}
\end{equation}
where $D_{max}$ is defined via $F(D) = 0$ for $D \geq D_{max}$.
In practice we set $D_{max}$ to the separation beyond which our results
no longer show a force significantly different from zero.  More extensive
simulations might lead to revised estimates of $D_{max}$, but integration of a weak, oscillatory force will have minimal influence on the results for the entropic potential near contact.  
(Note that the uncertainty in $D_{max}$ has no influence on our calculation of the
barrier height $\Delta V$, as this involves integrating the force from contact to the 
first separation (well below $D\!=\!1$), at which $F=0$.
To evaluate V(D) we form piece-wise polynomial fits to the force data.
\vspace{1em}

{\bf IV Results}
\vspace{1em}

We begin by comparing our simulation results for the wall-sphere scaled 
force profiles with the predictions of AO and HNCP theories,  
for diameter ratio $\xi \!=\! 5$ (Fig. 3) and 10 (Fig. 4).  We see that even at low density, 
AO theory underestimates the force 
at contact, and that this worsens with increasing density. The repulsive peak 
near $D=1$ grows more prominent with increasing density, and is of course 
absent from the AO prediction.  HNCP theory, by contrast, gives a very good 
account of the force at and near contact, and reproduces 
the detailed structure of the force profile, except for 
underestimating the repulsive peak at higher densities.  
(Since HNCP theory is exact only through ${\cal O}(\rho^3)$, it is not 
surprising that it grows less accurate with increasing density.)
Note as well
the very close agreement between the force profiles for the two diameter ratios.
The entropic force evidently increases more rapidly than $\rho k_B T$ 
(the ideal-gas pressure), as AO theory would
have it.  It is therefore of interest to check whether the entropic force increases as
the actual pressure in the microsphere fluid, and to this end we compare in Fig. 5
the scaled force at contact $f^*_{ws} (0) \equiv F(0)/(2 \pi R \rho k_B T)$ and $\tilde{f}(0) \equiv F(0)/(2\pi R p)$
where $p$ is the pressure as given by the Carnahan-Starling equation \cite{CARNAHAN}.
We see that while the entropic force grows much more rapidly than the density,
it does not grow nearly as rapidly as the pressure.  
It is also evident from this plot that $F(0) \propto R$ to very high precision, and
that $lim_{\rho \rightarrow 0} f^*_{ws} (0) \simeq 1$, as expected.
The force profiles for a pair of macrospheres, shown in Figs. 6 and 7, parallel 
the pattern observed in the wall-sphere simulations.    
Again we observe generally good agreement between HNCP theory and simulations, with 
the largest relative discrepancy appearing at the largest volume fraction.
Here there is a greater discrepancy between theory and
simulation regarding the force at contact; HNCP overstates the magnitude of the force by about
1/3, for diameter ratio 5.

Having data for both sphere-sphere and wall-sphere interactions, albeit at somewhat different
densities, affords us the opportunity of making a direct test of the Derjaguin approximation,
$f^*_{ss} = f^*_{ws}$.  Accordingly we interpolate the $\xi =10$ wall-sphere force to
volume fraction $\eta = 0.229$, and compare the resulting scaled force with that for
the sphere-sphere case at the same diameter ratio and volume fraction (see Fig. 8).
Since the scaled forces show no evidence of a significant, systematic difference,
we conclude that the Derjaguin approximation is reliable to within statistical
uncertainty (5\% or less over most of the range).

In Fig. 9 we plot the entropic potential derived from the wall-sphere simulations for $\xi=10$.
With increasing density, the height of the repulsive barrier increases relative to the depth of the minimum at contact, so that for $\eta = 0.3$ the potential
difference between the barrier and the second well is about $6 k_B T$ for this particle size.  Fig. 10 shows that the HNCP prediction for $V(D)$  (for $\xi=10$ and
$\eta = 0.3$) is in good agreement with simulation (it is even more so at lower densities), with most of the discrepancy arising from its underestimate of the repulsive barrier. (On the
other hand, it seems reasonable to ascribe the disagreement in the range $D = 2$ --- 2.5 to scatter in the simulation data.)

Of particular interest are the entropic potential at contact, and the barrier height $\Delta V$ for
leaving the wall, since the latter is the principal determinant of the time required for
escape from the wall \cite{KAPLAN94a}.  We compare theoretical predictions for
these parameters against simulation results in Table I (wall-sphere) 
and Table II (sphere-sphere).  We estimate the
relative uncertainty in the simulation results for $V(0)$ as
5\%, and that in $\Delta V$ as 3\%.
Evidently there is no significant difference between simulation and HNCP regarding
the wall-sphere potential at contact.  For sphere-sphere interactions the
situation is less clear: for $\xi = 5$ HNCP overestimates $|V(0)|$, 
but for $\xi = 10$ there is good agreement 
between the HNCP predictions for $|V(0)|$ and $\Delta V$ and simulation results.  
It seems reasonable to expect HNCP to yield reliable predictions for the 
sphere-sphere entropic potential for $\xi \geq 10$.

The good agreement
between simulation and HNCP theory is encouraging, and is of course consistent with 
the close match of the force profiles.  More surprising is the accuracy of
the AO prediction for $V(0)$, based, as it is, on a force profile that is quite different from
simulation.  The agreement between AO theory and simulation is somewhat
fortuitous, as is revealed by a consideration of the barrier height, $\Delta V$.
The repulsive peak observed in simulations, and predicted by HNCP theory, leads to
a barrier considerably in excess of the contact potential.  But AO theory is
insensitive to this distinction, predicting, as it does, a purely attractive force of
range $2r$.  (It is perhaps worth remarking that despite the presence of a repulsive
barrier, the entropic interaction is attractive overall, i.e., it makes a net negative
contribution to the second virial coefficient between macrospheres.)

\begin{table}
 \begin{center}
  \caption{Theory and simulation values for the wall-sphere entropic potential at
   contact.  $V$ is given in units of $k_B T$. }
  \label{tab1}
  \begin{tabular}{llllll}
$ \eta $ &  $V_{AO} $ & $V_{HNCP}$ & $\Delta V_{HNCP}$ & $V_{MC}  $ & $\Delta V_{MC}$   \\ 
\multicolumn{6} {c} {$ \xi = 5$}  \\
0.10   & -1.60  & -1.64 &  1.77 & -1.64  & 1.77  \\
0.20   & -3.20  & -3.36 &  3.93 & -3.48  & 3.99   \\
0.30   & -4.80  & -5.05 &  6.62 & -5.22  & 6.86   \\ 
\multicolumn{6} {c} {$ \xi = 10$}  \\
0.10   & -3.10  & -3.16 &  3.42 & -3.22  & 3.51  \\
0.20   & -6.20  & -6.40 & 7.57  & -6.43  & 7.88   \\
0.30   & -9.30  & -9.43 & 12.71 & -9.21  & 13.20         \\ 
  \end{tabular}
 \end{center}
\end{table}

\begin{table}
 \begin{center}
  \caption{Theory and simulation values for the sphere-sphere entropic potential at
   contact.  $V$ is given in units of $k_B T$. }
  \label{tab2}
  \begin{tabular}{llllll}
$ \eta $ &  $V_{AO} $ & $V_{HNCP}$ & $\Delta V_{HNCP}$ & $V_{MC}  $ & $\Delta V_{MC}$   \\ 
 \multicolumn{6} {c} {$ \xi = 5$}  \\
0.116   & -0.99  & -1.03 &  1.12 & -0.91  & 1.00  \\
0.229   & -1.95  & -2.13 &  2.51 & -1.84  & 2.17   \\
0.341   & -2.90  & -3.30 &  4.36 & -2.89  & 3.69    \\ 
 \multicolumn{6} {c} {$ \xi = 10$}  \\
0.116   & -1.86  & -1.91 &  2.09 & -1.73  &  2.04  \\
0.229   & -3.66  & -3.86 &  4.65 & -3.69  &  4.54   \\
0.341   & -5.46  & -5.64 &  7.96 & -5.70  &  8.24    \\ 
  \end{tabular}
 \end{center}
\end{table}
\vspace{1em}

{\bf V Discussion}
\vspace{1em}

Recently Kaplan, Facheux and Libchaber devised an ingenious method for
measuring the entropic force between a macrosphere and a hard wall in a
binary colloid mixture \cite{KAPLAN94a}.  The screening length in their suspension 
of polystyrene spheres is sufficiently short that the interactions are well-aproximated by hard-sphere potentials.  The depth of the potential well for a macrosphere at the
wall is determined from an estimate of the escape time, the latter being inferred from time-series
of the particle's {\em transverse} Brownian motion, which reflects proximity to the wall
through a sharp change in the diffusion coefficient.  The reported barrier heights range from
about 1.5 $k_BT$ at $\eta = 0.1$ to about $3 k_B T$ at $\eta = 0.3$.  These results indicate
a much weaker force than we observe in our simulations; taking our $\xi=10$ results and
scaling them up to the experimental value of $\xi=28.6$, we would expect $\Delta V / k_BT
\simeq 10 - 38 $ for this range of volume fractions.  We are unable to explain this order-of-magnitude discrepancy between our theory and simulations, on one hand, and the experimentally measured barriers on the other.  Given the close agreement between the HNCP
approximation and simulation, it appears unlikely that either is subject to massive error.  Other
possibilities are significant departures from hard-sphere interactions, and/or a
substantial correction to the effective diameters of the spheres, and difficulties in estimating
the attempt frequency $\tau_D^{-1}$ \cite{KAPLAN94a}.

Very recently, in an experiment on a binary colloid, Dinsmore, Yodh and Pine measured  
the entropic barrier encountered by a macrosphere near a wall, in the vicinity of an 
edge \cite{DINSMORE}.  (The reduction in overlap of exclusion regions as the 
macrosphere approaches the edge leads to a free energy increase relative to its being 
at or near contact with a planar wall.)
In this case experiment is in reasonable accord with the corresponding AO prediction.
While we have not studied this geometry, our results for sphere-sphere and 
wall-sphere interactions lead us to
expect that AO theory would be in fair agreement with simulations and HNCP theory
regarding the potential at a corner as well.  Further refinements in experimental
technique should render detailed comparisons of experimental and theoretical force 
profiles feasible.  

At the level of the two-body effective interaction considered in this work, the 
macrosphere fluid is characterized by a hard core and a (mainly) attractive short-range tail
whose depth increases with the microsphere density.  If the entropic effects of microsphere packing and interstitial configurations can be ignored, one might hope to predict the
phase diagram of the macrosphere system on the basis of the entropic potential found from
simulation or HNCP theory.  We defer this task to future work, but note that our present study
yields predictions that should be amenable to an experimental test.  Using the HNCP
sphere-sphere entropic potential, we calculate the reduced second virial coefficient
$B^*_2 \equiv B_2/b_0$, where $b_0 = 16 \pi R^3 /3$ is the hard-core second virial
coefficient, and find, for $\xi = 10$, values of 0.72, 0.00, and -1.78 for volume fractions
0.116, 0.229, and 0.341, respectively.  In other words, our theory predicts that the macrosphere
fluid with diameter ratio 10 is at its Boyle point for a microsphere volume fraction
of about 0.23; for $\xi = 20$, $B^*_2$ vanishes when $\eta \simeq 0.12$.  
Osmotic pressure or compressibility measurments at low macrosphere concentrations
should permit verification of our predictions.

In summary, we find that the HNCP theory of Attard and Patey yields quite 
accurate predictions for
the entropic potential between a macrosphere and a hard wall, and between 
a pair of macrospheres, induced by a hard microsphere fluid.
For wall-sphere interactions we observe no significant disagreement between HNCP 
theory and simulation data regarding the potential at contact or the barrier height.
For sphere-sphere interactions we observe a modest tendency of HNCP
theory to overestimate the strength of the interaction for diameter ratio $\xi\!=\!5$,
but theory and simulation are in very good agreement for $\xi\!=\!10$.
One may therefore be confident in applying HNCP to systems with larger diameter 
ratios, and that the Derjaguin approximation, which we used to derive the wall-sphere
entropic potential, given the sphere-sphere predictions of HNCP theory, is
reliable in this context.  Direct comparison of wall-sphere and sphere-sphere
force data also supports the latter conclusion.
(The good agreement for both diameter ratios also indicates that our simulations
are not subject to significant finite-size corrections.) The accuracy of the HNCP
force profile does show signs of breaking down at higher densities, where the
repulsive peak is not faithfully reproduced.  While
the simple AO theory yields severly inaccurate force profiles, its prediction of the
entropic potential at contact is, perhaps accidentally, reasonably good.  
\vspace{1em}

\begin{flushleft}
{\bf Acknowledgments}
\end{flushleft}
\vspace{1em}

We thank Peter Kaplan for helpful discussions.

\newpage

\noindent{\bf Figure Captions}
\vspace{1em}

\noindent FIG. 1.  
Microsphere density $\rho$ vs. distance from wall $x$ in a simulation with a single
macrosphere; $R=5$, $D=2.5$, $H=22$, and bulk volume fraction $\eta =0.3$.
\vspace{1em}

\noindent FIG. 2. 
Extrapolation of the shell integral $I(s)$ to contact ($s=5.5$).  The data are for a
single macrosphere; $R=5$, $D=1$, and bulk volume fraction $\eta =0.3$.  The solid
line is a least-squares cubic fit to the ten points nearest contact.
\vspace{1em}

\noindent FIG. 3. Entropic force profiles between a hard macrosphere and a hard wall,
in a background hard microsphere fluid, for diameter ratio $\xi = 5$.  Symbols: simulation
results; solid curve: HNCP theory; dashed line: AO theory.  Upper panel: volume fraction
$\eta = 0.1$; middle: $\eta = 0.2$; bottom: $\eta = 0.3$.
\vspace{1em}

\noindent FIG. 4. Same as Fig. 3 but for $\xi = 10$.
\vspace{1em}

\noindent FIG. 5. Scaled entropic force at contact versus background fluid density.
$+$: $|f^*(0)|$, $\xi =5$; $\circ $: $|f^*(0)|$, $\xi =10$; 
$\times$: $|\tilde{f}(0)|$, $\xi =5$; $\Box$: $|\tilde{f}(0)|$, $\xi =10$. 
\vspace{1em}

\noindent FIG. 6. Entropic force profiles between a pair of hard macrospheres
in a background hard microsphere fluid, for diameter ratio $\xi = 5$.  Symbols: simulation
results; solid curve: HNCP theory; dashed line: AO theory.  Upper panel: volume fraction
$\eta = 0.116$; middle: $\eta = 0.229$; bottom: $\eta = 0.341$.
\vspace{1em}

\noindent FIG. 7. Same as Fig. 6 but for $\xi = 10$.
\vspace{1em}

\noindent FIG. 8. Test of the Derjaguin approximation for $\xi=10$ and $\eta=0.229$.
$\bullet : f^*_{ws}$; $\circ : f^*_{ss}$.
\vspace{1em}

\noindent FIG. 9. Entropic potential from simulations of a macrosphere at a hard wall, $\xi = 10$, 
for volume fractions $\eta = 0.1$ (lowest peak), 0.2, and 0.3 (highest peak).
\vspace{1em}

\noindent FIG. 10. Wall-sphere entropic potential, $\xi = 10$, $\eta=0.3$. Solid line:
simulation; dashed line: HNCP theory.
\vspace{1em}

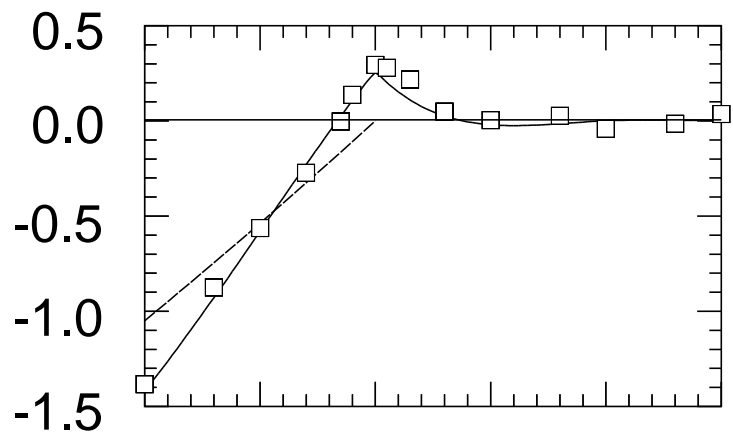
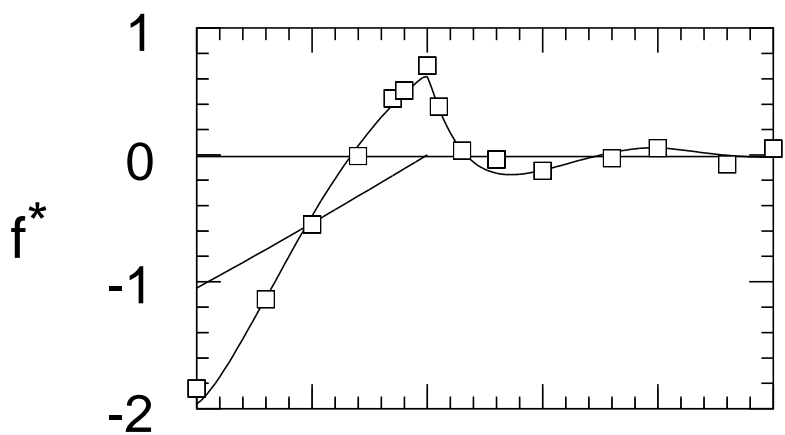
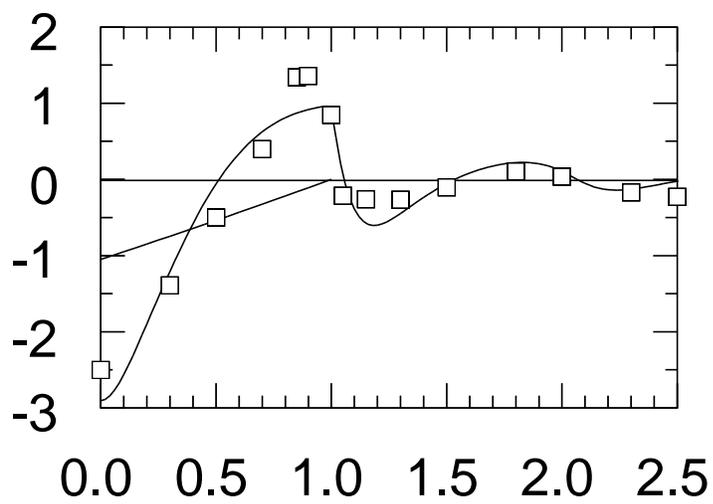
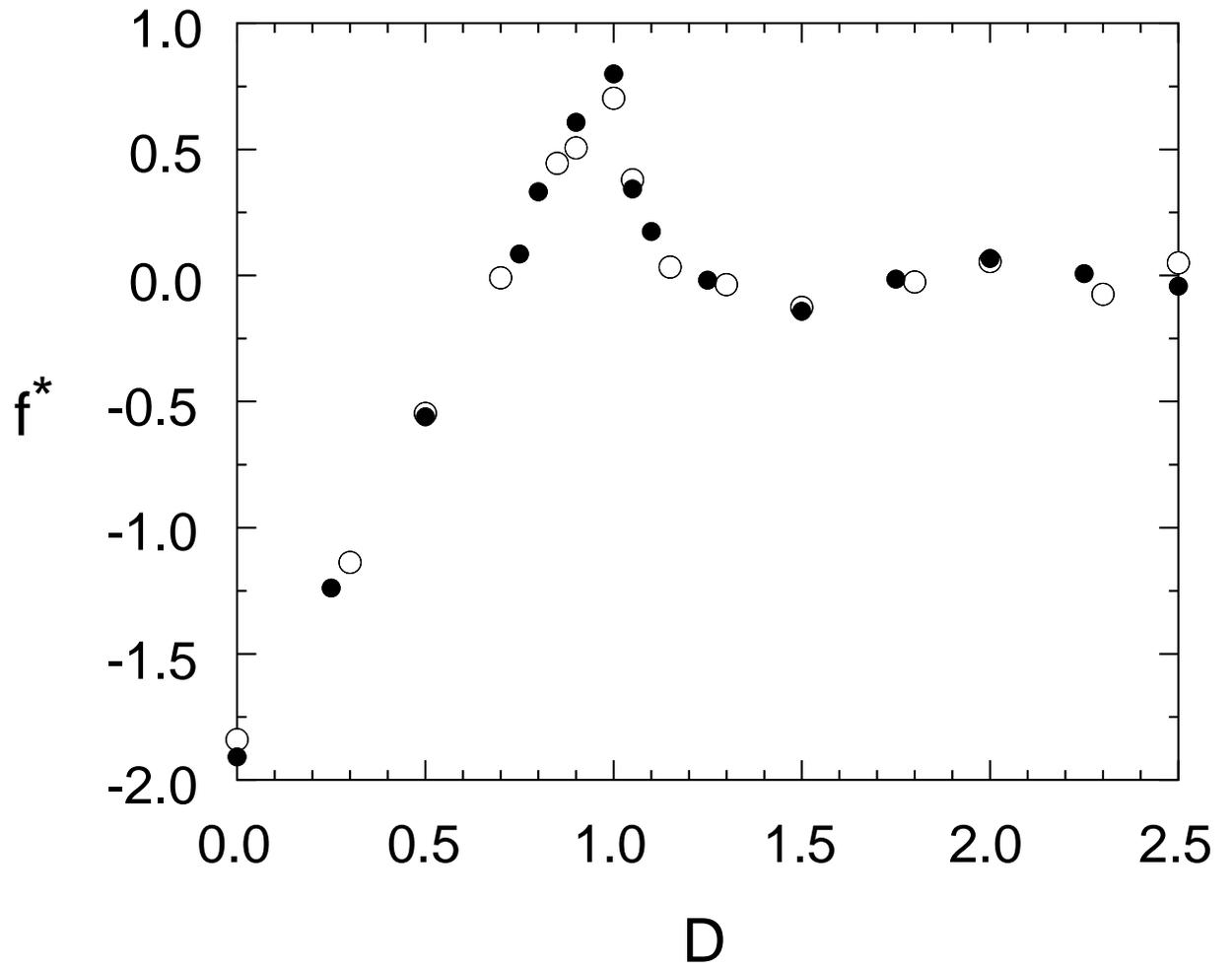
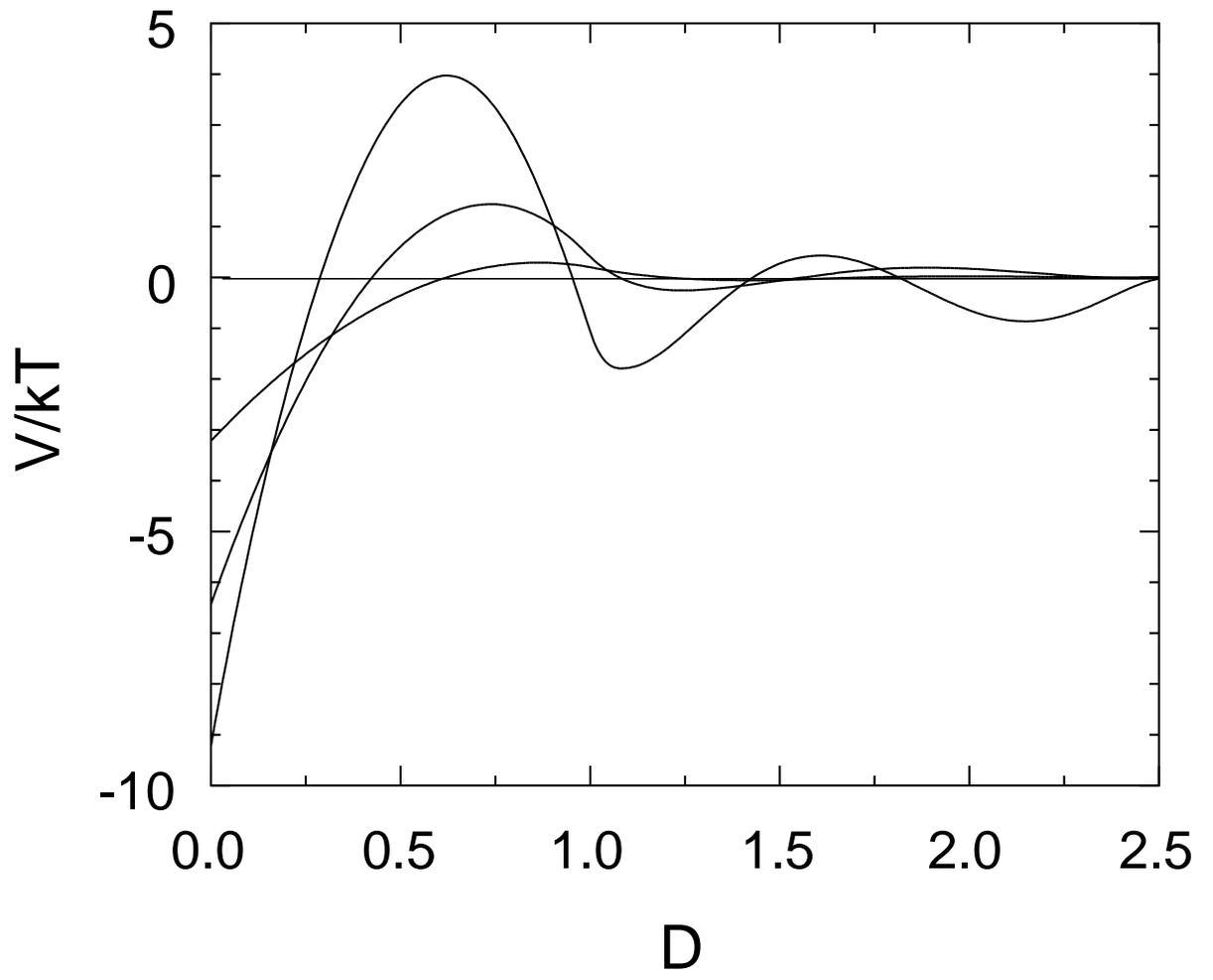
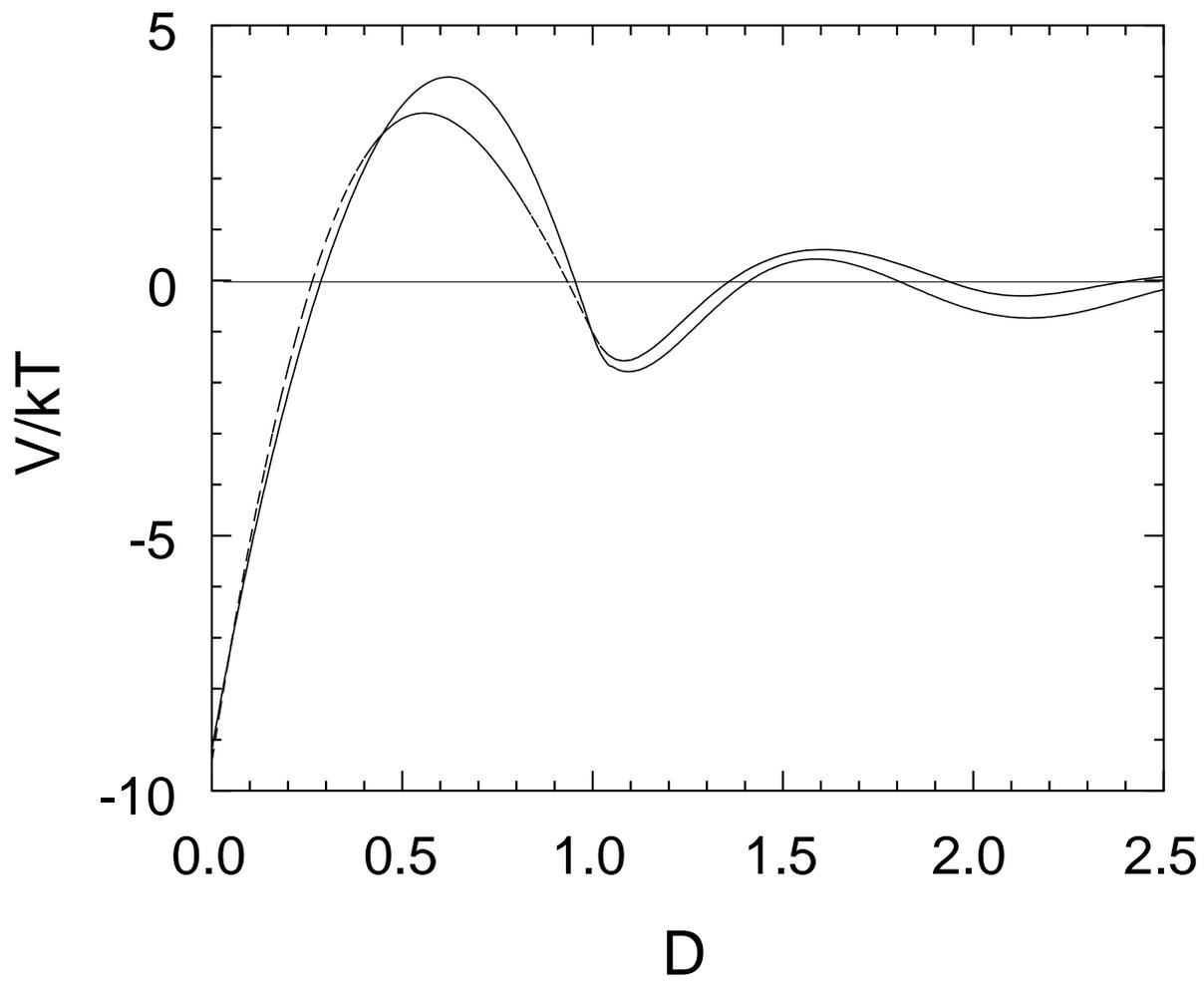
\end{document}